# µDopplerTag: CNN-Based Drone Recognition via Cooperative Micro-Doppler Tagging

O. Yerushalimov, D. Vovchuk, A. Glam, and P. Ginzburg

*Abstract*— The rapid deployment of drones poses significant challenges for airspace management, security, and surveillance. Current detection and classification technologies, including cameras, LiDAR, and conventional radar systems, often struggle to reliably identify and differentiate drones, especially those of similar models, under diverse environmental conditions and at extended ranges. Moreover, low radar cross sections and clutter further complicate accurate drone identification. To address these limitations, we propose a novel drone classification method based on artificial micro-Doppler signatures encoded by resonant electromagnetic stickers attached to drone blades. These tags generate distinctive, configuration-specific radar returns, enabling robust identification. We develop a tailored convolutional neural network (CNN) capable of processing raw radar signals, achieving high classification accuracy. Extensive experiments were conducted both in anechoic chambers with 43 tag configurations and outdoors under realistic flight trajectories and noise conditions. Dimensionality reduction techniques, including Principal Component Analysis (PCA) and Uniform Manifold Approximation and Projection (UMAP), provided insight into code separability and robustness. Our results demonstrate reliable drone classification performance at signal-to-noise ratios as low as 7 dB, indicating the feasibility of long-range detection with advanced surveillance radar systems. Preliminary range estimations indicate potential operational distances of several kilometers, suitable for critical applications such as airport airspace monitoring. The integration of electromagnetic tagging with machine learning enables scalable and efficient drone identification, paving the way for enhanced aerial traffic management and security in increasingly congested airspaces.

## I. Introduction

Controlling and monitoring airborne traffic is becoming an increasingly important challenge with the anticipated large-scale deployment of drones. These small autonomous vehicles are expected to carry out various logistical missions, including deliveries, agricultural management, surveillance, and more [1]. Despite the obvious advantage of this approach, it comes with severe security challenges and loopholes related to the safe management and operation of airspace. This is particularly relevant in urban areas, where high population density, numerous obstacles, and diverse traffic participants introduce additional complexities. Several regulatory measures have been implemented to address these issues and enhance safety, including remote identification (RID), which requires drones to report their location in real time. This approach relies on continuously broadcasting geolocation data and is prone to potential failures despite its accuracy and efficiency. To address these challenges and provide an additional layer of safety, monitoring solutions that do not rely on any broadcast signal from the drone must be deployed. These systems must be capable of detecting drones and effectively managing airspace, even under challenging conditions.

In applications requiring reliable object detection and identification, especially in dynamic and challenging environments, cameras and LiDAR systems face significant limitations [2]. Cameras, while capable of capturing detailed visual information, are highly susceptible to environmental conditions such as fog, rain, snow, and low light, which can obscure or distort the captured imagery. This issue is particularly pronounced in low-light or nighttime scenarios where cameras without infrared capabilities struggle to perform effectively. Similarly, LiDAR is sensitive to adverse weather, and its real-time data processing requires significant computational resources, adding to deployment complexity. Of particular concern is that both cameras and LiDAR have difficulty distinguishing between objects with identical appearances, such as drones from the same vendor, which further limits their effectiveness in critical identification tasks. These challenges emphasize the need for alternative or complementary sensing technologies that can overcome these limitations [3].

Radar detection systems often compromise price, performance, and reliability. They can operate under both day and night conditions and are less affected by weather, allowing for the detection of targets over long distances [4]. The integration of advanced front-end hardware and digital signal processing algorithms enables efficient monitoring and tracking of distant targets and their classifications. Small drones are relatively new airborne targets that pose a challenge to radar systems due to their inherently small radar cross sections (RCS). Additionally, because their size is comparable to birds, reliable detection requires the use of advanced classifiers, which in turn demand monitoring at relatively high signal-to-noise ratios (SNRs), leading to an increased risk of false alarms [5]. The analysis of micro-Doppler signatures, which capture the frequency shifts caused by small, though complex movements within a target, such as the flapping of wings or rotating blades, has emerged as a promising approach for improving target detection and classification [6], [7], [8], [9]. While attempts have been made to harness its full potential, this remains an open field of research with much to explore. Recently, inspired by the success of deep learning techniques, significant efforts have been directed towards applying these methods to micro-Doppler, unlocking unique advantages like enhanced target identification and improved classification accuracy [10], [11], [12], [13].

*Our Identification System.* Considering the need for a reliable radar-based system for monitoring small drones, we propose a passive electromagnetic approach with several essential features. This system will enable the identification of

O. Yerushalimov, D. Vovchuk, A. Glam, and P. Ginzburg are with the School of Electrical Engineering, Tel Aviv University, Tel Aviv, Israel. *(Corresponding author: Dmytro Vovchuk,* e-mail: dimavovchuk@gmail.com*)*.

small drones from long distances and facilitate their classification with the assistance of a neural network. To enable the passive detection scheme and account for severe time-evolving clutter, as found in urban environments, several engineering parameters must be carefully considered and precisely designed.

(i) First, the radar cross-section (RCS) of a drone has to be enhanced. Typical values for the DJI Mini 2 are approximately 10 cm², considering a typical S- or X-band surveillance radar (with spectral ranges of 2-4 GHz and 8-12 GHz, respectively). While these values enable the drone to be detected from several kilometers in open-sky conditions, low-altitude flights and flights against backgrounds such as walls or trees make the drone barely distinguishable from the surrounding clutter. Furthermore, applying Doppler filtering is essential, as it limits detection to moving drones while excluding those that are hovering or flying in directions tangential to the radar antenna. However, enhancing the RCS is not the only requirement, as background clutter scatters more than the target. Therefore, spectral features must be introduced into the radar echo to allow for detection and classification.

(ii) The second parameter for passively increasing radar visibility is to introduce time variation to the RCS, effectively creating a blinking target. In this case, the time-dependent RCS will be achieved through the rotating blades of the drone, which will be equipped with efficient, lightweight stickers that do not compromise aerodynamic conditions. These stickers significantly enhance the micro-Doppler signature of the plastic blades, which naturally generates a weak micro-Doppler due to the low scattering efficiency of those thin and electromagnetically transparent structures. With strong micro-Doppler signatures, machine-learning techniques can be effectively applied for recognition.

(iii) The third requirement for the tagging system is the potential for code diversity, as drones from the same manufacturer may produce very similar and barely distinguishable signatures. Therefore, each drone should be equipped with a unique electromagnetic structure that generates a distinct micro-Doppler signature. Considering varying flight conditions, potential orientations of the drone relative to the radar antenna, and the presence of clutter, the diversity of codes must be comprehensively analyzed. The coding distance between signatures should be carefully designed to create an optimal set, ensuring high recognition accuracy with minimal false detections.

The concept of our *μDopplerTag* solution complies with all the demands outlined above. It is worth noting that the basic electromagnetic demonstration of this concept was previously reported by us [14], [15]. However, the aspects of code diversity, an efficient target recognition system based on a convolutional neural network (CNN), and evaluation of the detection algorithm applied on nontrivial flight trajectories were not addressed. Here, we conduct intensive data collection both indoors and outdoors across a wide range of different tag configurations and develop a new CNN to ensure reliable detection and classification between tagged drones.

The report is organized as follows: first, the electromagnetic aspects of scattering are discussed, followed by a demonstration of the data collection setup, which serves as the foundation for machine learning analysis. Before the classification stage, the signals are analyzed using classical spectrogram-based methods to identify the main micro-Doppler fingerprints. While the spectrographic approach offers valuable physical insights into the data, it leads to a significant loss of information when analyzed using image-based algorithms. To overcome this limitation, we have developed a CNN capable of processing raw radar data at the baseband level (I/Q data, sampled at a high rate), which is represented in a complex-valued form using the Short-Time Fast Fourier Transform (STFT) for pre-filtering.

## II. ELECTROMAGNETIC SYSTEM AND DATA COLLECTION

The interrogation process is conducted using a custom-made continuous-wave (CW) radar with high Doppler resolution, sufficient to perform an accurate micro-Doppler analysis, as shown below. This system consists of a PNA (Precision Network Analyzer) connected to a pair of antennas - one for transmission (Tx) and the other for reception (Rx). To improve electromagnetic isolation between the transmitter (Tx) and receiver (Rx), the antennas were placed 1 m apart, resulting in approximately 60 dB suppression of mutual coupling. Given that the target was located at distances greater than 10 m, this configuration can still be treated as effectively monostatic. A low-noise amplifier (LNA) is placed after the Rx port to boost the received signal, as illustrated in Fig. 1(a). The radar frequency is 3.35 GHz. Fig. 1(b) and Fig. 1(c) show photographs corresponding to the indoor and outdoor data collection procedures, respectively. The differences between the electromagnetically isolated anechoic chamber, with no reflections, and outdoor conditions are evident. Since the drone operates close to the ground, its radar returns are affected by ground reflections and static clutter.

The experimental procedure follows the methodology that we developed in [14], [15]. The radar carrier frequency was selected to be in the S-band, with the μDopplerTag designed to resonate at 3.35 GHz. This frequency corresponds to a wavelength of 10 cm; thus, a half-wavelength dipole would be 5 cm across. However, since the stickers are intended to be affixed to the blades of a small drone, which are typically around 5 cm in size, the resonator must be miniaturized to fit within this footprint. Among the various existing solutions, we favor meandering as it is easy to prototype and implement [16]. The meander is fabricated from adhesive foil and precisely cut using a milling machine (LPKF ProtoMat E33). The typical layout of the μDopplerTag, as attached to a blade, is shown in Fig. 1(d), along with the corresponding dimensions, which are presented in Fig.1(e)    . The core principle behind the time-dependent backscattering of the rotating dipole is the polarization mismatch. Considering the incident wave's polarization in the rotating plane of the rotor, the backscattering is maximal when the dipole is aligned with the polarization and minimized when the dipole and the electric field vectors are mutually orthogonal. This results in intensity modulation of the backscattered wave at a frequency corresponding to twice the rotation rate of the rotor [17], [18], [19]. Considering additional precession of the rotating shaft and varying angles of attack of the drone relative to the interrogating antenna, richer spectral signatures emerge, complicating detection. All these factors, which affect detection reliability, will be addressed and resolved using a neural network, as will become evident later.

It is worth noting that metalizing the entire blade does not improve the electromagnetic response, as the RCS of a structure that is not tuned to resonance remains lower. This variant also offers no manufacturing benefits.

Fig. 1f shows a photograph of the drone, with several blades tagged with meander structures while others remain untouched. This procedure establishes the tagging space, where a tagged blade is considered '1' and an untagged blade is '0'. Although the coding space theoretically consists of 8 bits, several of those binary codes may produce nearly identical electromagnetic signatures due to symmetries, rendering them redundant. Some redundant configurations, such as swapping the blades on the left and right sides, can be identified immediately, as the radar lacks sufficient resolution to distinguish between rotors in range and angular bins. However, other redundancies may be less obvious, as the geometrical considerations are non-trivial due to the differing operational behavior of the forward and rear blades. Additionally, diagonal blades rotate in opposite directions. The rotation is highly synchronized but can be influenced by various flight conditions such as wind, acceleration, and other external factors. While these complexities may seem to pose significant challenges to detection reliability, the developed CNN effectively handles them. The differences between tag codes will be systematically addressed by performing dimensionality reduction using two approaches, namely Principal Component Analysis (PCA) and Uniform Manifold Approximation and Projection (UMAP).

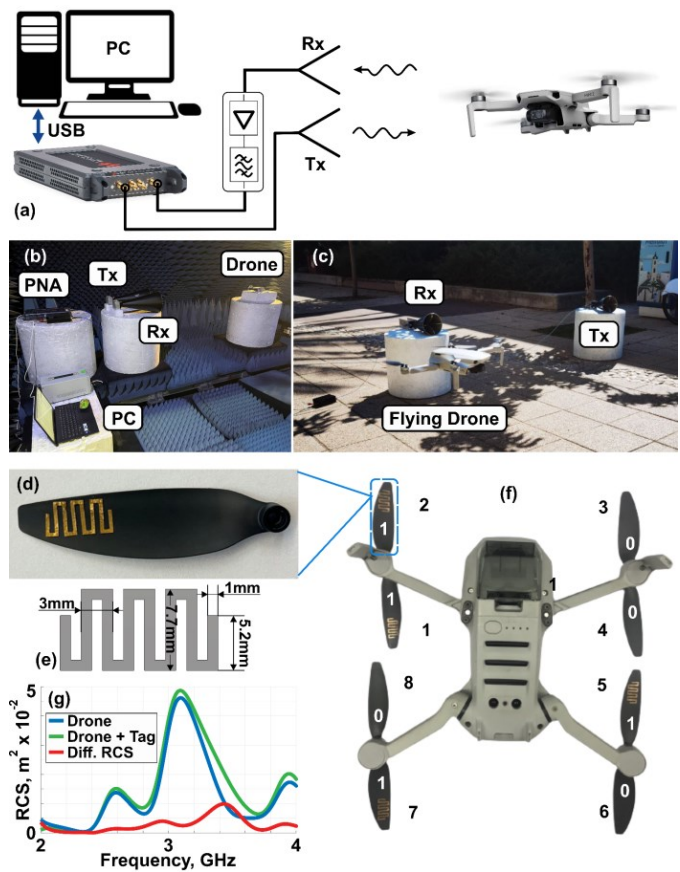

Fig. 1. (a) Block diagram of the custom-made continuous wave (CW) radar. (b) Photograph of the indoor experiment conducted in an anechoic chamber. (c) Photograph of the outdoor experiment. (d) Photograph of a drone blade with a meandered sticker attached (e) with depicted geometrical characteristics . (f) Photograph of a drone, where a tagged blade is considered as '1' in the code space and an untagged blade as '0'. (g) Experimentally measured radar cross sections (RCS) of a drone with a single tagged blade. The green curve represents the case where the tagged blade is aligned with the polarization of the incident wave; the blue curve corresponds to the rotor rotated by 90°(polarization mismatch); and the red curve shows the differential RCS, demonstrating a peak precisely at the designed operating frequency of the CW radar (3.35 GHz).

The SNR of micro-Doppler signatures critically influences the maximum achievable detection range. Although polymer-based rotor blades inherently produce micro-Doppler returns, their RCS is generally low, resulting in weak signals that limit reliable target classification at extended distances. Typically, such signatures are detectable only up to approximately 1 kilometer, even with advanced radar systems. The resonant stickers introduced here significantly enhance the electromagnetic response, thereby increasing the micro-Doppler SNR and extending the effective detection range. Fig. 1(g) presents the RCS of a drone equipped with a single tag. All the RCS measurements were performed in a certified chamber, using a brass disk as the calibration target. The drone is statically positioned in an anechoic chamber, facing the antenna within the plane of incidence. In the first experiment, the tagged blade on a drone is aligned with the incident wave polarization, while in the second, the blade is rotated by 90°, creating a polarization mismatch. The RCS spectra for both cases are shown in Fig. 1(g), along with their difference. The obtained differential RCS is of the same order of magnitude as that of the entire drone, effectively making the micro-Doppler signature (blinking target) as detectable as the drone itself. In other words, if a radar can detect a drone at 5 km (as claimed by a few vendors), it can potentially recognize the artificially induced micro-Doppler signature from a similar distance. To ensure reliable detection and enable tag recognition, a neural network is required.

III. DATA ASSESSMENT AND PRE-PROCESSING USING PHYSICS MODEL

*Indoor vs Outdoor Data*

Before analyzing the raw data and applying machine learning, it is essential to acknowledge that the data can be interpreted physically, providing meaningful insights. The indoor data was collected by measuring a hovering drone in an anechoic chamber. In this controlled environment, background reflections from walls were suppressed to typical levels of -50 to -40 dB, which effectively results in data with a very high SNR. Moreover, the absence of wind, acceleration, or other disturbances ensures synchronization in the rotational velocity of the propeller blades. As a result, the micro-Doppler signatures appear as distinct, spectrally narrow horizontal lines in the spectrograms. The comparison between indoor and outdoor data, obtained using the same drone, is presented in Fig. 2. For visualization purposes, a single antenna is shown in the inset, whereas Fig. 1 presents the complete experimental setup. For the sake of comparison, two codes, as indicated in the plots, are evaluated. The upper row corresponds to the measurements taken in the anechoic chamber, while the bottom row shows the results from the outdoor environment. The plots show spectrograms over 10-second intervals, which provide a sufficient duration for investigating radar targets. Detailed information on sampling and other parameters will follow. The purpose of this comparison is to demonstrate that, although the outdoor data is noisier and the peaks are broadened, it still preserves the underlying structure. This suggests that machine learning techniques for classification have a good chance of succeeding despite the flight conditions.

*Outdoor Data with Nontrivial Flight Paths*

In the case of linear motion, acceleration, or movement along non-trivial paths, the micro-Doppler signal is superimposed with the Doppler shift originating from the main return, which is associated with the motion of the drone's center of mass. Additionally, acceleration can cause abrupt changes in the rotational speed of the propeller blades, and the rear and front blades may rotate at different velocities. In this set of experiments, the outdoor wind was negligible. A set of spectrograms presented in Fig. 3 illustrates typical scenarios obtained using different flight paths and codes. In all plots, the lower highlighted frequency corresponds to the main Doppler shift. The motion trajectories are schematically depicted at the top of each plot. For example, Doppler nulling is observed when the drone changes its radial velocity, for instance, transitioning from approaching the radar to moving away. In such cases, the instantaneous radial velocity becomes zero at a certain time frame, resulting in Doppler nulling. Very low Doppler shifts are also observed during zig-zag motion. These behaviors can be identified in the spectrograms and correlated with the corresponding motion paths.

become dissimilar. As a result, the initially overlapping spectral signatures split into distinct pairs.

All drone flights were performed in the area that was covered by the radar's antennas; therefore, we excluded a situation in which the Doppler and micro-Doppler signatures cannot be detected. In the outdoor experiment, the antenna's half-power beamwidth was ~ 20° (gain 10dB), and the maximum range was 20 meters - the Drone flew within this area. The drone's flight altitude was 1.5 meters, and the CW radar operated at a frequency of 3.35 GHz.

## IV. CODE RECOGNITION WITH NEURAL NETWORK

Micro-Doppler analysis of low SNR poses distinct complexities in signal processing [20], [21]. These challenges include the limited availability of data due to the rarity and complexity of collecting specific signatures, difficulties in isolating these effects from noisy backgrounds, and the high variability in signals caused by minor changes in the target or environmental conditions, all of which complicate the analysis. All these factors contribute to errors and the need for significant computational power for real-time processing. Integrating micro-Doppler signatures with neural network algorithms, combining radar technology and machine learning, has gained substantial attention in recent years [22], [23]. Furthermore, deep learning techniques, which provide robust tools for extracting complex patterns from noisy data, highlight the potential of convolutional neural networks (CNNs) to adapt to the high variability inherent in micro-Doppler radar signals. CNNs have been applied to classify micro-Doppler signatures from both synthetic and real-world datasets, demonstrating superior accuracy compared to traditional machine learning approaches [12], [24].

The diversity of codes and the detection strategy will be assessed in two steps. The first step involves extensive analysis of indoor data, with synthetic noise superimposed. Due to the controlled laboratory environment and ease of handling, multiple codes can be evaluated during this phase. The distances between the codes will be quantified, and the CNN outputs will be correlated with the visual data. The second step focuses on assessing outdoor data, where real-world conditions apply, though fewer realizations will be tested.

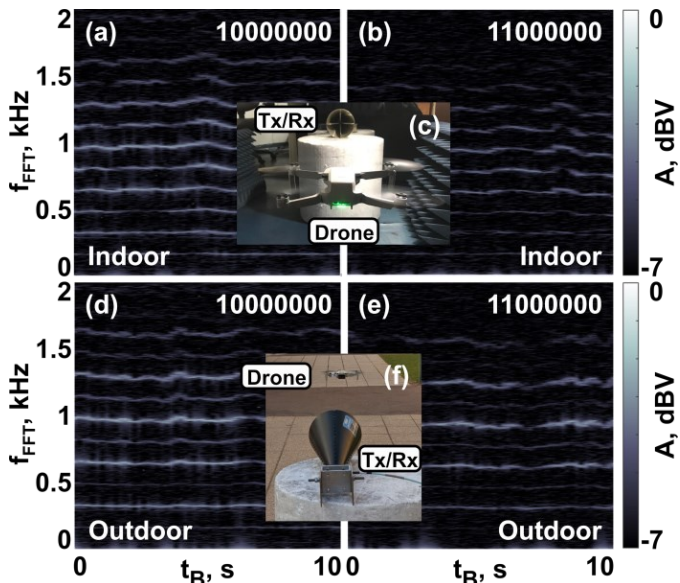

Fig. 2. Comparison of indoor and outdoor spectrographic data to evaluate the impact of flight conditions. (a) and (b) correspond to the indoor measurements, while (c) shows a photograph from the laboratory. (d) and (e) correspond to the outdoor measurements, with (f) showing a representative photograph from the outdoor experiment. The data is presented as spectrograms and Fourier transforms over the entire observation interval, with the corresponding codes indicated in the legends.

Considering the micro-Doppler signatures, the following observations can be made. First, the closer the drone is to the radar, the stronger the signatures appear. This is expected due to the increased SNR. When the drone accelerates, it typically tilts, causing the velocities of the rear and forward blades to

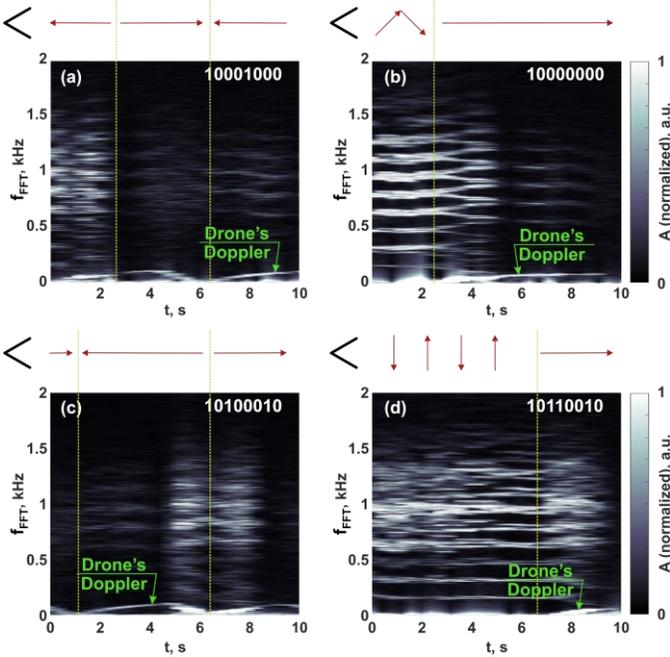

Fig. 3. Spectrograms for different representative codes and motion paths, as indicated on the plots. Changes in the trajectories are indicated by vertical dashed lines. The green arrows indicate the direction of the drone's flights related to the radar's antennas. The lower branch corresponds to the Doppler signatures, as marked.

*Indoor data, the structure*

Records from a stationary drone in the anechoic chamber were taken over 10 seconds, consisting of 50,001 samples. Each tag class has three records, which are further divided into 1.54-second segments with overlaps for enhancing the generalization of Doppler-based tag classifiers on unseen data. Considering the rotational speed of approximately 200 Hz, several seconds of data encompass multiple blade revolutions, making it sufficiently reliable and coherent time interval for processing.

There are 43 classes considered: 00000000, 00000010, 00000011, 00001000, 00001010, 00001011, 00001100, 00001110, 00001111, 00100000, 00101010, 00101111, 00110000, 00111111, 10000000, 10000010, 10000011, 10001000, 10001010, 10001100, 10001111, 10100000, 10100010, 10101000, 10101010, 10101111, 10110000, 10111111, 11000000, 11000010, 11000011, 11001000, 11001100, 11001111, 11100000, 11101111, 11110000, 11110010, 11111000, 11111010, 11111011, 11111110, 11111111 among all the possible $2^8$ combinations. These codes do not account for trivial redundancies, which are eliminated based on symmetry considerations. However, some non-trivial aspects related to propeller synchronization require further analysis, and the 43 combinations mentioned are not claimed to be exhaustive. A larger number of codes were not investigated owing to handling limitations.

*Outdoor data, the structure*

The outdoor measurements were conducted in an area measuring 20 meters by 5 meters, where arbitrary flying movements of a drone were recorded. The tests were conducted by initiating zig-zag trajectories, linear velocities, and accelerations to encompass most typical flight paths (see Fig. 3). Each recording lasted 15 seconds and consisted of 90,001 samples. For each class, there were 10 records. Before processing, the data was segmented into 2-second fragments with overlaps to augment the dataset. The dataset comprises 7 classes: 00100010, 10100010, 10110010, 11100000, 11110000, 10000000, and 11000000. The limited number of classes is solely related to the logistics of the outdoor measurement. These codes were selected based on separation within a dimensionality reduction map.

*Convolutional Neural Network Design*

In contrast to previous studies [15], the spectrograms are not treated as images, where complex-valued data is lost. Instead, the input to the CNN is a 2D matrix (frequency vs. time) of complex-valued numbers. This approach ensures that no information is lost; instead, the data is simply restructured, making it more suitable for the neural network's processing. At the next stage, the matrix is unwrapped into a 1D vector, transforming it into a single-channel input. This approach enables the development of lightweight architecture. The CNN consists of three convolutional layers: the first layer applies 32 filters, followed by 64 filters in the second layer and 128 filters in the third, all with a 3x3 kernel size and ReLU activations. Max pooling with a 2x2 window is applied after each convolutional layer to reduce spatial dimensions, and dropout is used for regularization. After the convolutional layers, the output is flattened and passed through a fully connected layer with 128 neurons and ReLU activation, followed by a dropout layer. The final fully connected layer outputs the class predictions. This architecture strikes a balance between feature extraction and regularization to prevent overfitting. The output of the CNN is reformatted into a confusion matrix, which is evaluated separately for indoor and outdoor scenarios. The radar's performance is analyzed using a synthetic noise approach applied to the real collected data. This assessment enables the estimation of detection probability and code recognition accuracy as functions of SNR, providing a basis for predicting the feasibility of long-range detection in practical scenarios.

Inputs to the CNN are single-channel spectrogram matrices. The backbone comprises three 2D convolutional layers (32/64/128 filters, 3×3 kernels, ReLU), each followed by 2×2 max-pooling and dropout (p = 0.5). The feature map is then flattened and passed through a fully connected layer (128 units, ReLU, dropout p = 0.5) and a final classifier. Training uses cross-entropy with Adam ($l_r$ = 0.001), a batch size of 32.

*Summary on the Dataset Generation*

Indoor data were collected with the drone hovering in an anechoic chamber. Each raw recording lasted 10 s, containing 50,001 I/Q samples. Every record was converted into STFT spectrogram windows of 1.54 s with overlap inside a single sequence. Resulting frames were stored as single-channel inputs to the CNN.

Outdoor data were obtained from free-flight trajectories with linear, zigzag, and accelerating motion within a 20 x 5-meter region. Each raw recording lasted 15s with 90,001 samples. Signals were segmented into windows of 2 seconds with a size of 400x59 and an overlap of 1 second within each sequence. A model with 7classes was trained.

Indoor and outdoor experiments were collected, processed, and trained separately. Although 7 outdoor codes formed a subset of 43 indoor variants, underlying recordings remained disjoint and were never mixed during training or evaluation.

*CNN applied to Indoor Data*

In this analysis, the data obtained in the anechoic chamber is considered to have an infinite SNR, which is a reasonable approximation since, typically, instrumental noise is negligible compared to environmental noise, which is suppressed in the chamber.

For outdoor experiments, classification used a 3-layer CNN with 32, 64, and 128 filters, 3×3 kernels, rectified linear units, and 2×2 max pooling after each block, followed by dropout with a probability of 0.5 and a fully connected layer of 128 that feeds a 7-class softmax output, trained on single-channel spectrograms with 400x59 windows. Optimization was performed using cross-entropy loss and the Adam algorithm with a learning rate of $1 \times 10^{-3}$ and a batch size of 32, without any pre-training. The dataset was divided randomly into 80% for training, 10% for validation, and 10% for testing. To study robustness to SNR, the convolutional model was first fitted on clean spectrogram windows, and during evaluation, the held-out test samples were degraded by adding zero-mean Gaussian perturbations scaled to target levels of 0, 5, 7, 9, and 13 dB, while the training sequences remained unchanged.

Fig. 4a shows the confusion matrix for a selected set of codes (which, as will become evident, are the best distinguishable), demonstrating almost perfect classification performance from high SNR conditions (with no synthetic noise added) down to a remarkably low 0 dB SNR level. Across this entire SNR range, the matrix remains diagonal, highlighting the exceptional classification accuracy. The confusion matrix for all 43 codes is then calculated, and the procedure is repeated for data with synthetic noise added. Due to their large size, these matrices are not presented here. The confusion matrices are subsequently post-processed to extract information regarding detection reliability. Fig. 4b summarizes the data, demonstrating that an overall accuracy of 99% is achieved at SNR levels above 9 dB. However, at 0 dB, the classification becomes unreliable. The values were extracted by evaluating all 43 classes and averaging the results over the entire sampling space. It is worth noting that typical SNRs required for a high classification radar accuracy generally fall within a range of 10 dB [25], although they are strongly dependent on the use case and implementation. While the averaged values provide an assessment of the entire set, they do not reveal which codes are robust and which are prone to misclassification.

Alongside the robust codes, there is a set that depends quite strongly on noise. The confusion matrix at four SNR levels for these codes is presented in Fig. 5. The physical interpretation in this case can be clearly drawn. For example, in the 5 dB SNR plot, the '10000011' code is misclassified as '11100000'. In this case, the forward and rear rotors, with both blades tagged, interchange. Since there is no resolution on the target, and the electromagnetic rescattering between the rotors is weak [18], the origin of the misdetection is quite evident.

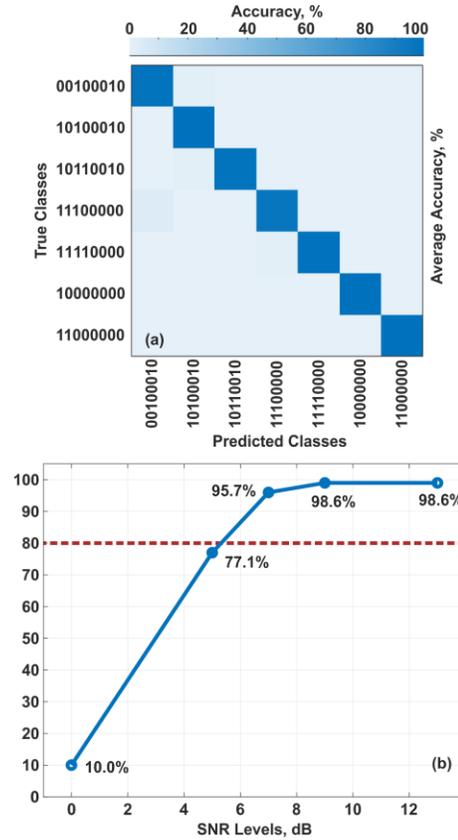

Fig. 4. Detection accuracy for indoor data. (a) Confusion matrix for a subset of codes robust to noise, showing a nearly diagonal matrix across five SNR levels, spanning from 9 dB to 0 dB. (b) Averaged accuracy for 43 codes as a function of SNR, obtained with synthetic noise.

To clarify the train validation test protocol and to rule out data leakage, indoor experiments employed a leave-one-sequence-out strategy across three recordings per code, where two files served for training and the remaining one was reserved for evaluation in each fold. During the SNR study, additive white Gaussian noise was injected only into held-out test windows at evaluation time, while network weights remained fixed across all SNR levels. The high accuracy reported in Fig. 4b, therefore, does not arise from trivial memorization. Each test frame originates from raw data streams that never contributed to parameter updates, due to the separation of recording levels. Moreover, indoor performance curves in Fig. 4b increase from roughly 10% at 0dB to about 99% above 9dB, and the evolution of confusions follows code symmetries illustrated in Fig. 5, which would be unlikely under a leaky protocol that simply replays training examples. Finally, the convolutional architecture remains compact, consisting of three blocks with 3x3 kernels, followed by a fully connected layer of size 128 and dropout with a probability of 0.5, which limits capacity and further discourages overfitting.

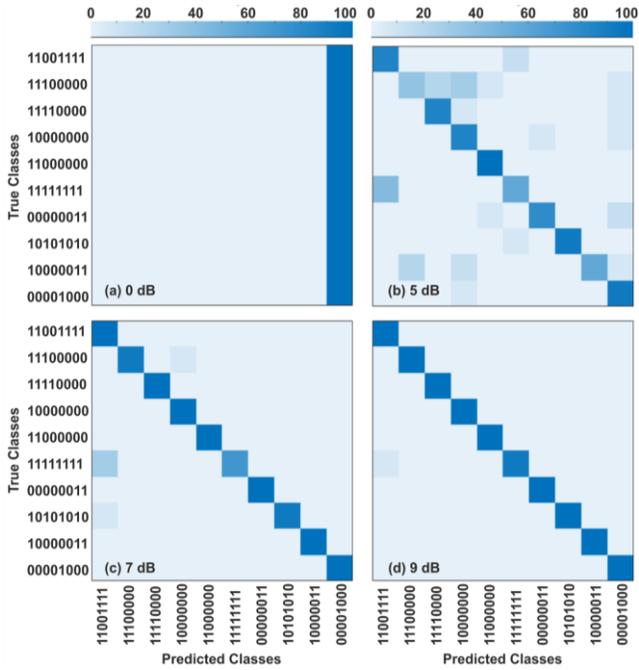

Fig. 5. Row-normalized confusion matrices for the 'difficult' indoor subset at 0, 5, 7, and 9 dB. Color denotes the percentage of samples per true class assigned to each predicted class (rows sum to 100%). The diagonal corresponds to per-class recall (TPR); off-diagonals are misclassification rates. Overall accuracy at each SNR is given above the panels and matches Fig. 4b.

While the confusion matrix provides numerical insight into the classification results, dimensionality reduction techniques offer a way to visually capture and interpret the relationships between codes in a lower-dimensional space.

The most commonly applied method is Principal Component Analysis (PCA). PCA transforms the data into a set of orthogonal components (principal components), ordered by the amount of variance they explain in the data. This allows for the reduction of the dataset's dimensions while retaining as much variability as possible. Typically, the first two principal components capture around 80-90% of the data variation in most cases. Fig. 6a shows the PCA embedding of spectrogram windows. For this visualization, each time-frequency magnitude map in dB was flattened into a single vector, then standardized with a zero mean and unit variance over the full window set. The resulting features were projected onto a 2D plane using this linear transform. Each marker represents the centroid of embedded windows for a single code, and the color encodes indoor accuracy without added noise. These representations were derived in an unsupervised manner and did not impact CNN training or evaluation, as labels were used only for averaging and coloring.

From the PCA plot, it is evident that several codes cluster together, indicating that under realistic SNR conditions, they may become indistinguishable, while others are clearly separated. To recap, the first principal component represents the most significant variance in the data, whereas the second component is of secondary importance.

To visualize this claim, spectrograms for three codes are shown in Figs. 6b, 6c, and 6d. For example, '10001000' and '10000010' (panels b and c) are physically very similar, differing only by a tag swapped between the two rear rotors. As a result, although these codes are somewhat distinguishable under very high SNR, they appear very close on the PCA map and, as seen in the confusion matrices in Fig. 5, become completely indistinguishable at lower SNR levels. On the other hand, '00110000' (panel d) is a physically different code, where one rotor is tagged with two stickers (one per blade), making it distinguishable from the others. This distinction is also apparent from the spectrograms, which are visually different. Note that the PCA plot itself does not state that both classes are perfectly recognized; this information is provided by the confusion matrix.

Another dimensionality reduction technique is Uniform Manifold Approximation and Projection (UMAP). Unlike PCA, which primarily focuses on maximizing variance, UMAP is designed to preserve the topological structure of the data, making it especially effective for high-dimensional datasets with complex, non-linear relationships. Applying UMAP to our dataset is particularly relevant because it retains both local and global relationships. Specifically, our data exhibits non-linear behavior, i.e., flipping a single bit (e.g., adding a sticker to a blade) can cause a dramatic change in the spectrogram. Therefore, UMAP is hypothesized to provide a more accurate and representative analysis in this context.

Fig. 7 presents a similar analysis to the one previously performed using PCA. However, this mapping reveals some remarkable behaviors. The three codes selected for analysis are highlighted, along with their corresponding spectrograms. While the codes in panels (c) and (d) are visually similar and consequently positioned close to each other on the map, the code in panel (b) is visually very different yet still appears within the same region. This suggests that these three codes form a cluster despite their spectral differences, which is counterintuitive. One possible explanation is that UMAP's reliance on preserving local and global topological structures can sometimes cause it to group data points that share subtle, high-dimensional features but appear distinct in lower-dimensional visualizations, such as spectrograms. Additionally, UMAP's results depend heavily on parameter choices (such as the number of neighbors and minimum distance), which can affect cluster formation and potentially lead to misleading proximity between dissimilar codes. Finally, noise or variability in the data might influence the embedding, causing unrelated codes to appear closer in the reduced space.

*CNN applied to Outdoor Data*

After revealing the most robust and distant codes by assessing a large number of classes indoors, the top 7 candidates were selected for the outdoor assessment. Specifically, the following binary sequences were chosen: 00100010, 10100010, 10110010, 11100000, 11110000, 10000000, and 11000000.

The outdoor data has been processed in a similar manner to the indoor data. However, the major difference lies in the non-trivial flight trajectories and the presence of severe ground clutter, as the flight altitude was only 1.5 meters. To assess the environmental impact, Gaussian noise was synthetically added to the data during the test stage. The radar chart presented in Fig. 8a simultaneously visualizes several key metrics: accuracy, precision, recall, F1 score, and false positive rate (FPR) across several SNR levels. Accuracy measures the overall correctness of predictions. Precision specifically reflects the proportion of predicted positive cases that are indeed correct. Recall captures how effectively the model identifies actual positive cases from the dataset. The F1 score provides a balanced combination of precision and recall, especially valuable when precision and recall differ significantly. In contrast, the false positive rate explicitly shows the proportion of cases wrongly identified as positive, highlighting the model's tendency toward incorrect predictions.

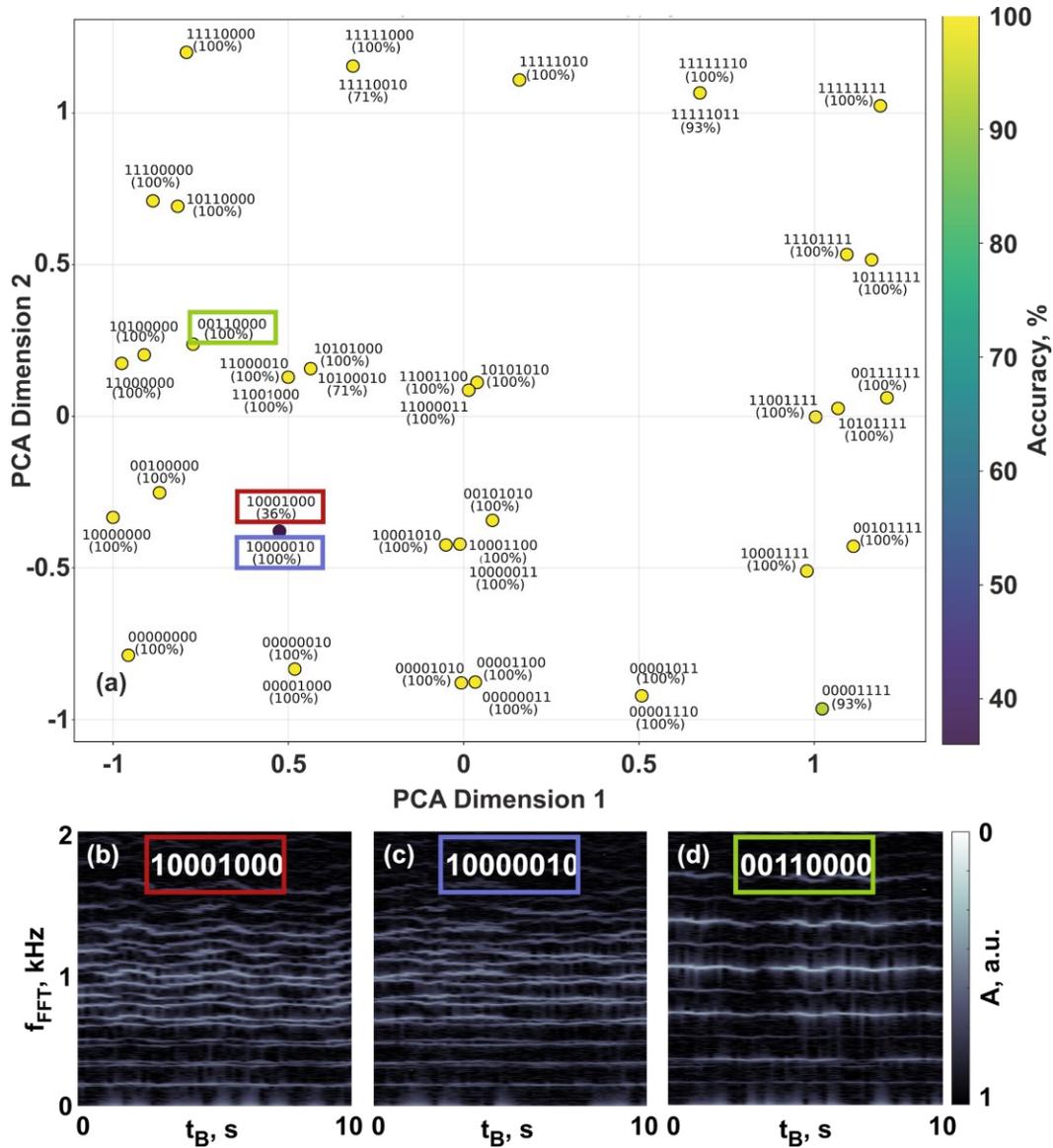

Fig. 6. Dimensionality reduction using Principal Component Analysis (PCA) to demonstrate code clustering and separation. (a) PCA map for 43 codes, with detection accuracy indicated by color (no noise added). The three codes selected for subsequent analysis are highlighted with colored frames. (b), (c), and (d) Spectrograms of the highlighted codes.

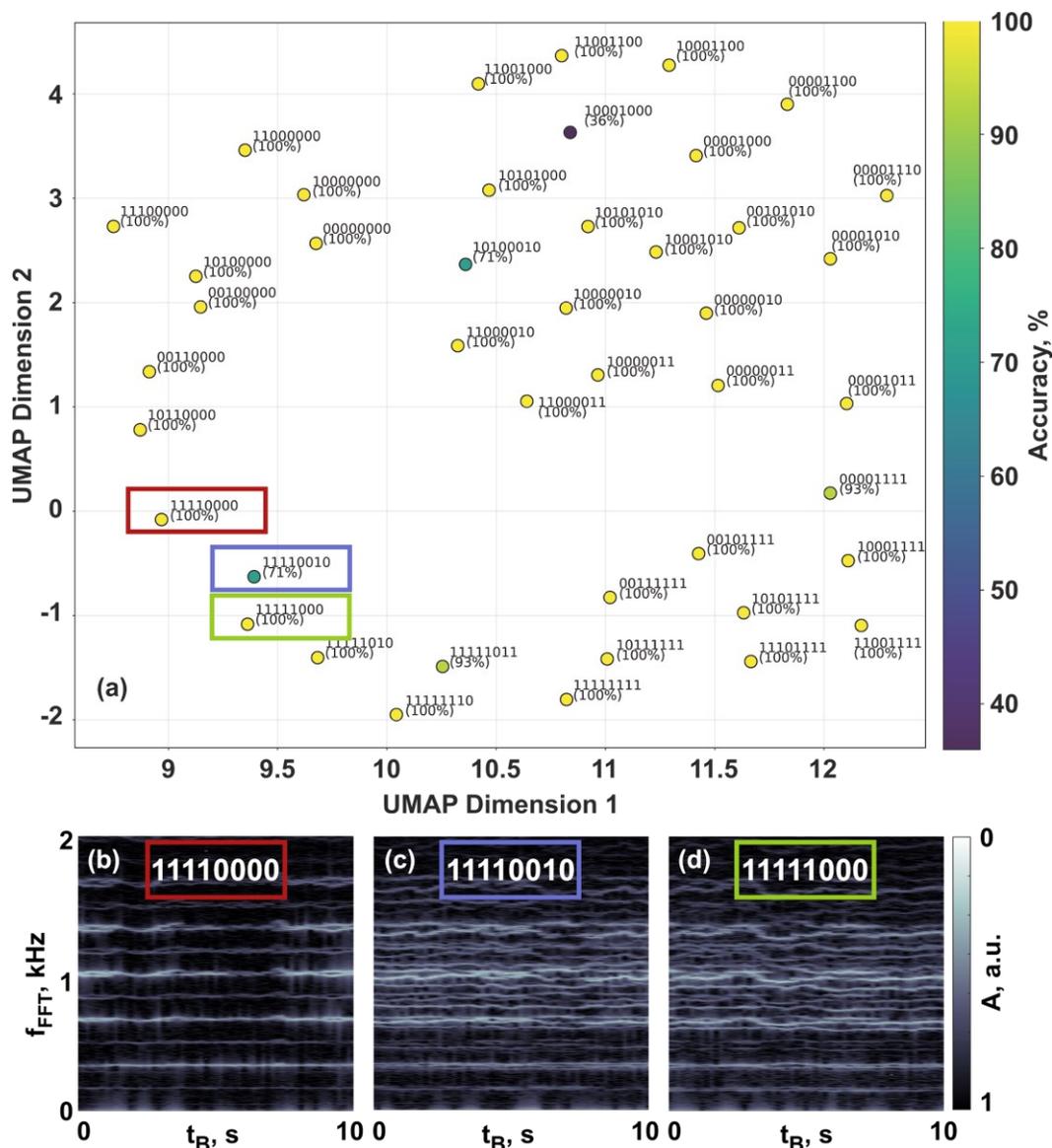

Fig. 7. Dimensionality reduction using Uniform Manifold Approximation and Projection (UMAP) to demonstrate code clustering and separation. (a) UMAP map for 43 codes, with detection accuracy indicated by color (no noise added). The three codes selected for subsequent analysis are highlighted with colored frames. (b), (c), and (d) Spectrograms of the highlighted codes.

Insights from this radar chart include the observation of a clear performance improvement across all metrics as SNR increases, with accuracy, precision, recall, and F1 score rising sharply, while the false positive rate decreases distinctly. The inverse trend of the false positive rate compared to other metrics explicitly emphasizes how the reduction in incorrect predictions aligns with the improvement in overall model reliability at higher SNR levels.

The data can be restructured differently, showing true positives (TP), false negatives (FN), true negatives (TN), and false positives (FP) as a function of SNR, as appears in Fig. 8b. Overall, as previously mentioned, a 7 dB SNR suggests accurate target classification. It is essential to note that such high accuracy is achieved by selecting codes that are well-separated on the PCA map.

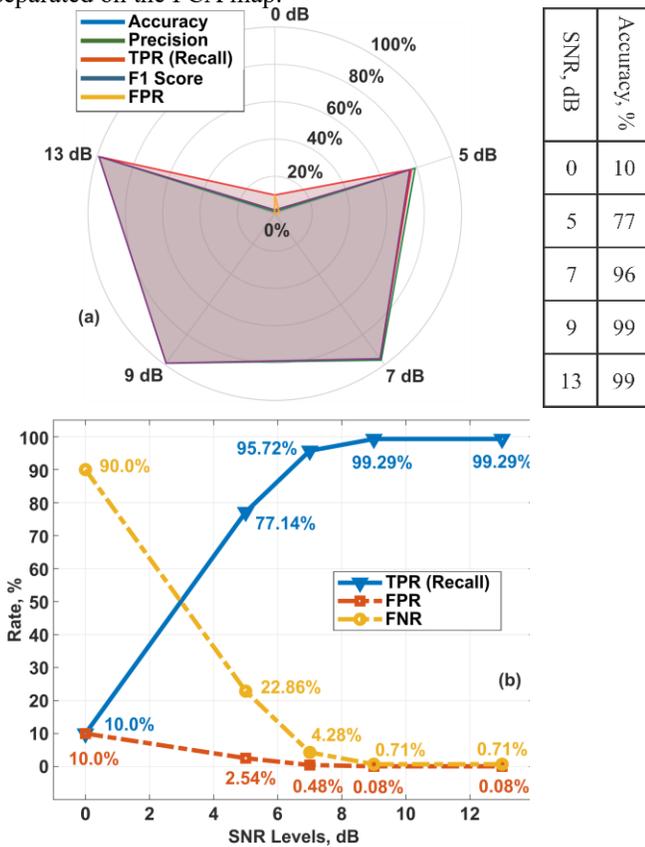

Fig. 8. Outdoor data classification performance. (a) Polar plot demonstrates the balance between 5 key parameters (accuracy, precision, recall, F1 score, and false positive rate across several SNR levels. Inset – table of classification accuracies. (b) True positive rate (TPR), false negative rate (FNR), true negative rate (TNR), and false positive rate (FPR) as a function of SNR.

## V. Conclusions

A novel approach to drone classification has been developed by designing artificial micro-Doppler signatures encoded through resonant electromagnetic stickers affixed to drone blades. These selectively attached stickers modulate the radar return in a configuration-specific manner, generating distinctive signatures observable at the radar's baseband signal. The proposed system integrates a tailored CNN architecture that demonstrates exceptional classification accuracy across both controlled indoor and realistic outdoor environments, maintaining robustness over a broad range of SNRs.

In the controlled environment of an anechoic chamber, 43 unique tag configurations were systematically measured under very high SNR conditions and analyzed using the CNN. The addition of synthetic noise enabled the exploration of code clustering behavior and the identification of the most distinguishable signatures. Classical dimensionality reduction via PCA provided an intuitive visualization of code separability, offering insights into which tag patterns are most resilient under harsh conditions.

Building upon these results, the seven most promising tag candidates were further evaluated in outdoor experiments featuring complex flight trajectories, including zig-zag maneuvers and accelerations. Despite the increased environmental complexity and lower SNR scenarios, the CNN maintained reliable detection performance down to SNR levels as low as 7 dB. This robustness demonstrates the practical feasibility of the method when deployed with high-grade air surveillance radar systems. This suggests that the *μDopplerTag system* can achieve detection and classification distances of several kilometers, addressing critical operational needs such as comprehensive drone monitoring over expansive airspaces, including major airports. The demonstrated integration of resonant electromagnetic tagging with advanced machine learning thus presents a scalable, efficient solution to emerging challenges in autonomous aerial traffic management and airspace security.

From a system perspective, several approaches exist for reporting drone identity and position. Regulatory solutions based on Remote ID require an onboard transmitter that broadcasts GNSS-assisted messages over Wi Fi or Bluetooth, with typical module costs on the order of tens of dollars per drone and full dependence on continuous power and radio integrity. Transponder-based schemes, such as ADS-B, offer larger ranges and integration with existing aviation infrastructure, but incur higher payload costs and weight, and are not mandated for small platforms in most jurisdictions. Purely non-cooperative surveillance based on micro-Doppler signatures avoids any modification of the airframe. However, classification performance degrades quickly with decreasing radar cross-section and clutter, and the decision logic remains sensitive to platform similarity. In contrast, the proposed resonant sticker represents a passive tag with negligible mass and a per-unit cost comparable to simple printed circuit elements, requiring no power and allowing for interrogation by standard radar hardware. This shift of complexity from onboard electronics to a low-cost electromagnetic pattern enables robust code separability at long range while keeping both infrastructure and per-drone expenses compatible with large-scale deployment.

## Appendix 1

Convolutional Neural Network Design:
Backbone (shared by indoor/outdoor).

- Conv block 1: Conv2D 3×3, 32 filters, ReLU → MaxPool 2×2 → Dropout p = 0.5.
Output map size: indoor 32×166×29; outdoor 32×200×29.
- Conv block 2: Conv2D 3×3, 64 filters, ReLU → MaxPool 2×2 → Dropout p = 0.5.
Output map size: indoor 64×83×14; outdoor 64×100×14.
- Conv block 3: Conv2D 3×3, 128 filters, ReLU → MaxPool 2×2 → Dropout p = 0.5.
Output map size: indoor 128×41×7; outdoor 128×50×7.
- Head. Flatten → FC-128, ReLU, Dropout p = 0.5 → FC-C (softmax over C classes; C = 43 indoor, C = 7 outdoor).

## Appendix 2

Confusion Matric for 43 classes:

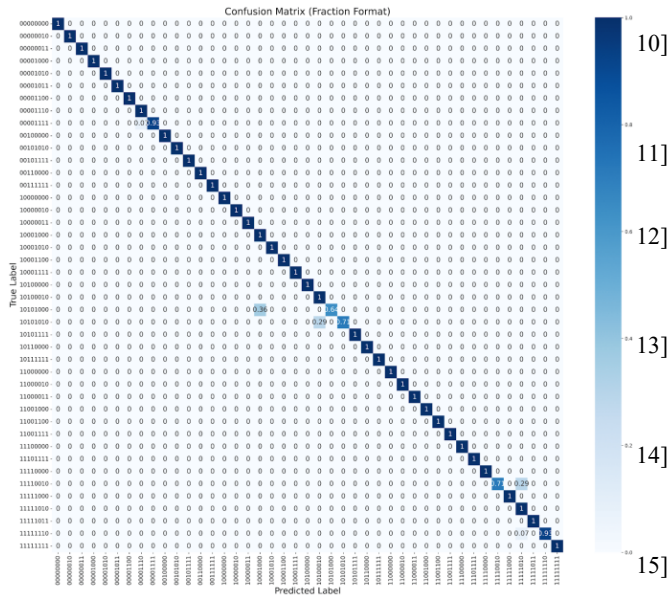